\documentclass[amsmath, amssymb, aps, pra, 12pt]{revtex4-2}
\usepackage{physics}
\usepackage{graphicx}% Include figure files
\usepackage[colorlinks=true,allcolors=blue]{hyperref}% add hypertext capabilities

\newcommand{\bk}[2]{\left< #1 \middle| #2 \right>}

\begin{document}

\title{Analytic level statistics of the quantum Rabi model}

\author{Zi-Min Li}
%\email{zimin.li@csu.edu.cn}

\author{Long-Jie Li}
\affiliation{Institute of Quantum Physics, Hunan Key Laboratory of Nanophotonics and Devices, School of Physics, Central South University, Changsha 410083, China}

\author{Germ\'an Sierra}
%\email{german.sierra@csic.es}
\affiliation{Instituto de F\'isica Te\'orica UAM-CSIC and Quantum Advance Center Research, Madrid, Spain}

\author{Murray T. Batchelor}
\email{murray.batchelor@anu.edu.au}
\affiliation{Mathematical Science Institute, Australian National University, Canberra ACT 2601, Australia}
\affiliation{Centre for Modern Physics, Chongqing University, Chongqing 400044, The People's Republic of China}

\begin{abstract}
We derive an analytic expression for the distribution of level spacings in the quantum Rabi model.
At high energies, the spectrum in a fixed parity sector takes the asymptotic form of a uniformly spaced ladder modulated by a decaying oscillation.
The alternating sign of this oscillation separates successive spacings into two branches.
For a finite window containing many oscillations, counting how often the two branches exceed a given spacing yields an analytic expression for the spacing density at leading order.
The oscillation amplitude at the lower end of the window sets the range of spacings, and its value at the upper end fixes the positions of two cusps that form twin peaks.
Spacing histograms from exact diagonalization agree with the analytic density, including its spacing range and twin peak positions.
At leading order, the same spectral analysis also yields an expression for the ratio of consecutive spacings, a statistic that requires no spectral unfolding.
The resulting ratio statistics agree closely with those from exact diagonalization.
For windows that continue to contain many oscillations, the spacing and ratio peaks move toward unity at higher energies as the spectrum approaches a uniformly spaced ladder.
The same counting approach can be explored in other systems with controlled spectral asymptotics, slowly varying spacing branches, and windows containing many oscillations.
\end{abstract}

\maketitle

\section{Introduction}
Level statistics characterize correlations in quantum spectra.
The standard contrast is between Poisson statistics, associated with uncorrelated spectra in integrable systems, and Wigner--Dyson statistics, associated with level repulsion in random-matrix ensembles, including the Gaussian orthogonal ensemble (GOE) \cite{Percival1973, Berry1977, Guhr_1998, Bohigas1984a}.
Many structured quantum systems nevertheless exhibit reproducible spectral correlations that follow neither limiting law.
An analytic spacing distribution connects these correlations to their spectral origin and complements comparisons with universal random-matrix curves.

The quantum Rabi model (QRM) is a minimal model of light-matter interaction \cite{Rabi_1936,Rabi_1937,Braak_2011,Xie_2017}.
It describes a two-level system coupled to a single bosonic mode and possesses a $\mathbb{Z}_2$ parity symmetry.
At high energies, the spectrum in each parity sector approaches a uniformly spaced ladder modulated by a decaying oscillation \cite{Schmutz1986,Feranchuk_1996,BoutetDeMonvel2021}.
A displaced oscillator basis provides a simple physical picture of the origin of this high-energy oscillation.

Previous numerical studies characterized QRM and spin-boson spacing statistics and used them as diagnostics of quantum chaos \cite{Graham_1984,Kus1985a,Graham1986,Lewenkopf_1991,Wang_2013}.
More recently, Ref.~\cite{ThiHoaiNguyen2024} used the asymptotics of large eigenvalues to show that the spacing distribution in each parity sector converges to a delta measure at unit spacing.
Related number-theoretic work characterized the fine structure of the full QRM spectrum, the parity types of almost all consecutive spacings, and an arcsine law for the rescaled eigenvalue deviations within each parity sector \cite{Rudnick2024}.
However, the spacing density has not been derived analytically for a finite index window at high energy within a fixed parity sector.

Here we derive an analytic expression for the spacing density over a finite index window at high energy before the limiting delta measure is reached.
Within one parity sector, the alternating sign of the decaying spectral oscillation separates consecutive spacings into two branches.
Counting the branch intervals above a spacing threshold yields the spacing density $P(s)$ at leading order.
The oscillation amplitude at the lower end of the window sets the support of $P(s)$, and its value at the upper end determines two square-root cusps that form twin peaks.
Spacing histograms from exact diagonalization agree with the analytic density, including its support and peak positions.
The same spectral analysis yields an expression for the ratio of consecutive spacings at leading order, a statistic that requires no spectral unfolding.
For windows that continue to sample many oscillations, shifting the index range upward moves the twin peaks toward $s=1$ and the ratio enhancement toward unity.

The remainder of this paper is organized as follows.
Section~II introduces the model and asymptotic spectrum.
Sections~III and~IV derive and benchmark the spacing density $P(s)$ and the ratio of consecutive spacings, respectively.
Section~V presents our conclusions.

\section{Model Hamiltonian and asymptotic spectrum}

\subsection{Model Hamiltonian}

The Hamiltonian of the quantum Rabi model (QRM) reads
\begin{equation}
    H_R = \omega a^\dagger a + \frac{\Delta}{2}\sigma_z + g \sigma_x (a^\dagger + a).
\end{equation}
Here $\omega$ is the bosonic mode frequency, $\Delta$ is the energy splitting of the two-level system, $g$ is their coupling strength, $a^\dagger$ and $a$ are bosonic creation and annihilation operators, and $\sigma_x$ and $\sigma_z$ are Pauli matrices.
The conserved parity operator is $\Pi=\sigma_z(-1)^{a^\dagger a}$, with eigenvalues $p=\pm1$.
We keep $\omega$ explicit while constructing the displaced oscillator basis and use dimensionless variables from the asymptotic spectrum onward.
The displaced basis makes the spectral oscillation and its alternating sign at fixed parity explicit.

\subsection{Displaced oscillators and the adiabatic approximation}

The adiabatic approximation (AA) gives a compact analytic description of the QRM spectrum in the displaced oscillator basis, especially when $\Delta/\omega\ll1$ \cite{Feranchuk_1996,Irish_2005,Li2021GAA,Li_2021}.
At $\Delta=0$, we work in the $\sigma_x$ basis, $\sigma_x\ket{\pm x}=\pm\ket{\pm x}$, and define the dimensionless displacement $q=g/\omega$.
In the two spin sectors, the Hamiltonian becomes
\begin{equation}\label{Hdisplaced}
H_\pm^{\Delta=0}
=\omega(a^\dagger\pm q)(a\pm q)-\frac{g^2}{\omega}.
\end{equation}
Using the displacement operator
\begin{equation}
\mathcal{D}(\pm q)=\exp[\mp q(a^\dagger-a)],
\end{equation}
the eigenstates and their common energy are
\begin{equation}\label{displacedStates}
\ket{n_\pm,\pm x}
=\mathcal{D}(\pm q)\ket n\otimes\ket{\pm x},
\qquad
E_n^{\Delta=0}
=n\omega-\frac{g^2}{\omega}.
\end{equation}
For each $n$, the two states are degenerate and correspond to oscillators displaced in opposite directions.

For $\Delta\ne0$, the term $\Delta\sigma_z/2$ couples the two displaced states, giving a tunneling interpretation of their splitting \cite{Li2021a}.
The AA retains only the coupling between states with the same $n$.
The tunneling matrix element between the two displaced states is
\begin{equation}\label{AAOmega}
\Omega_n^\text{AA}
=\frac{\Delta}{2}\bk{n_-}{n_+}
=\frac{\Delta}{2}e^{-2q^2}L_n\left(4q^2\right),
\end{equation}
where $L_n$ is the standard Laguerre polynomial.

Parity exchanges the two displaced states and contributes the oscillator factor $(-1)^n$.
For a given parity eigenvalue $p=\pm1$, the parity eigenstate therefore alternates between the symmetric and antisymmetric combinations as $n$ changes, giving
\begin{equation}\label{EAAparity}
E_{n,p}^\text{AA}
=n\omega-\frac{g^2}{\omega}+p(-1)^n\Omega_n^\text{AA}.
\end{equation}
For fixed $g$, the Laguerre factor makes the tunneling matrix element $\Omega_n^\text{AA}$ vary in magnitude and change sign as $n$ increases.
The additional factor $(-1)^n$ changes sign at every step in $n$, producing the level-to-level alternation within each parity sector.
The eigenvalue formula at large $n$ introduced in the next subsection provides the quantitative input for the statistical analysis and does not rely on the AA restriction to equal oscillator indices.

\subsection{Asymptotic eigenvalues at large level index}

We henceforth measure energies and couplings in units of $\omega$ and set $\omega=1$.
Let $\lambda^p(n)$ denote the ordered eigenvalue with index $n$ in the parity sector $p$, where $p=+1$ denotes even parity and $p=-1$ denotes odd parity.
For $g>0$ and large $n$, these eigenvalues obey \cite{Feranchuk_1996,BoutetDeMonvel2021}
\begin{equation}\label{asymptoticEigenvalues}
\lambda^p(n)=n-g^{2}+p(-1)^{n} \frac{\Delta}{2 \sqrt{2 \pi g}} n^{-1 / 4} \cos \left(4 g \sqrt{n}-\frac{\pi}{4}\right)+O\left(n^{-1 / 2+\epsilon}\right).
\end{equation}
Here $\epsilon>0$ is an arbitrarily small exponent in the remainder estimate, not a parameter of the Hamiltonian.
The estimate is taken as $n\to\infty$ at fixed $g>0$ and fixed $\Delta$. Its implicit constant can depend on $g$, $\Delta$, and $\epsilon$.
Choosing $0<\epsilon<1/4$ makes the remainder smaller than the $n^{-1/4}$ envelope of the oscillatory correction.
Theorem~1.1 of Ref.~\cite{BoutetDeMonvel2021} applies at any fixed $g>0$ and fixed real $\Delta$, with no small-splitting restriction.

\begin{figure}[!t]
\centering
	\includegraphics[width=0.88\linewidth]{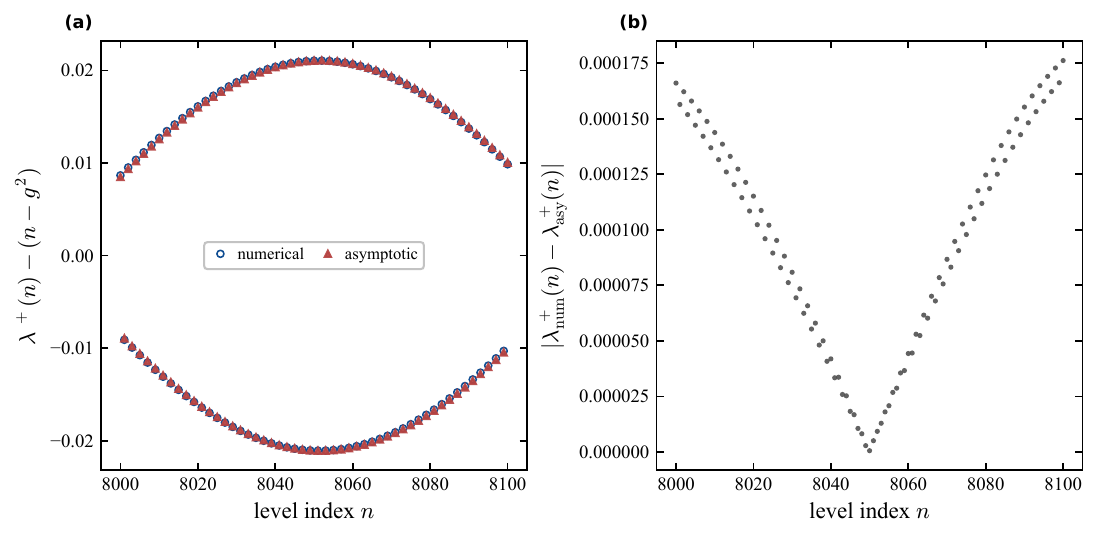}
	\caption{Spectrum benchmark at high energy in the even-parity sector for $g=\Delta=1$.
	(a) Numerical eigenvalues (open blue circles) and the asymptotic values from Eq.~(\ref{asymptoticEigenvalues}) (filled red triangles), after subtracting the leading ladder $n-g^2$ in the window $n\in[8000,8100]$.
	(b) Absolute difference between the numerical and asymptotic eigenvalues at each index in the same window.}
	\label{fig:spectrum_window}
\end{figure}

We benchmark Eq.~(\ref{asymptoticEigenvalues}) against exact diagonalization of the tridiagonal QRM Hamiltonian in each parity sector.
We truncate each parity block at $N_{\mathrm{cut}}=110000$.
For $g=\Delta=1$ and the largest index window analyzed below, increasing the cutoff to $120000$ changes the consecutive spacings by at most $8.8\times10^{-10}$, confirming numerical convergence for the central benchmark.
After subtracting the leading ladder $n-g^2$, Fig.~\ref{fig:spectrum_window}(a) shows that the asymptotic formula reproduces the alternation between adjacent levels and the slow residual oscillation.

The maximum error in panel (b) is $1.762\times10^{-4}$, less than one percent of the local oscillation amplitude.
Writing $\theta_n=4g\sqrt n-\pi/4$, the phase increment is $\theta_{n+1}-\theta_n=4g/(\sqrt{n+1}+\sqrt n)\sim2g/\sqrt n$.
Thus, away from the zeros of the cosine, its sign persists over neighboring indices while $(-1)^n$ changes sign at each step.
This produces the local alternation within a parity sector. Changing the parity eigenvalue $p$ reverses the oscillatory correction.
This alternating correction generates the two spacing branches analyzed in Section~III.

\section{Analytic spacing statistics}

\subsection{Spacing branches and envelope}

Let $\lambda(n)$ denote the ordered eigenvalues in one parity sector.
We choose even parity for definiteness and write $\lambda(n)=\lambda^+(n)$.
At leading order, odd parity only interchanges the assignment of indices to the two spacing branches.
The spacing is
\begin{equation}\label{sn}
s(n)=\lambda(n+1)-\lambda(n).
\end{equation}
Using Eq.~(\ref{asymptoticEigenvalues}), the spacing has the asymptotic form
\begin{equation}\label{snapprox}
s(n)=1+(-1)^{n+1}\frac{\Delta}{2 \sqrt{2 \pi g}}\left[\frac{\cos \left(4 g \sqrt{n+1}-\frac{\pi}{4}\right)}{(n+1)^{1/4}}+\frac{\cos \left(4 g \sqrt{n}-\frac{\pi}{4}\right)}{n^{1/4}}\right]+O\left(n^{\epsilon-1/2}\right).
\end{equation}
For $n\gg1$ with $2g/\sqrt{n}\ll1$, the oscillation amplitude and phase vary slowly over one index step, and the leading spacing is
\begin{equation}\label{spacingLeading}
s(n)\simeq1+(-1)^{n+1}r n^{-1/4}\cos\left(a\sqrt{n}-\frac{\pi}{4}\right),
\end{equation}
where
\begin{equation}
r=\frac{\Delta}{\sqrt{2\pi g}},\qquad a=4g.
\end{equation}
We take $g>0$ and $\Delta>0$, for which $r$ is a positive amplitude coefficient.
A negative $\Delta$ interchanges the branch labels. The envelope, support, and ratio depend on $|r|$.
At $\Delta=0$, the displaced-oscillator spectrum in Eq.~(\ref{displacedStates}) of Sec.~II.B gives exactly unit spacings in either parity sector.

The leading sequence alternates between the two functions
\begin{equation}\label{spm}
s_{\pm}(n)=1\pm r n^{-1/4}\cos\left(a\sqrt{n}-\frac{\pi}{4}\right).
\end{equation}
In the even-parity sector, odd $n$ samples $s_+(n)$ and even $n$ samples $s_-(n)$. 
The odd-parity sector reverses this assignment.
Thus the superscript on $\lambda^\pm(n)$ labels parity, whereas the subscript on $s_\pm(n)$ labels spacing branches within one parity sector.
In the continuum approximation below, the integer index is treated as a continuous variable and the discrete count is replaced by an integral over many oscillations.
The resulting continuous, normalized approximation to the empirical spacing distribution is called the leading continuum spacing density.
Changing parity leaves this leading density invariant.
Both branches lie inside the shrinking envelope
\begin{equation}
1-rn^{-1/4}\leq s_{\pm}(n)\leq1+rn^{-1/4}.
\end{equation}

\begin{figure}[!t]
\centering
	\includegraphics[width=\linewidth]{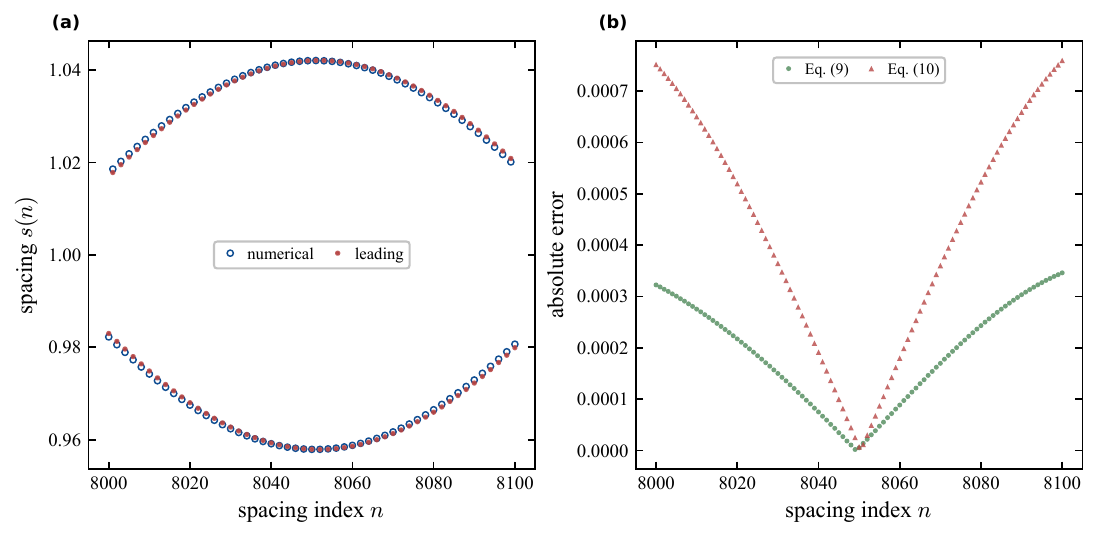}
	\caption{Spacing benchmark at high energy in the even-parity sector for $g=\Delta=1$ and $n\in[8000,8100]$.
	(a) Numerical spacings (open blue circles) and the leading values from Eq.~(\ref{spacingLeading}) (filled red circles).
	(b) Absolute errors from Eq.~(\ref{snapprox}) (green circles) and Eq.~(\ref{spacingLeading}) (red triangles).}
	\label{fig:spacing_window}
\end{figure}

Figure~\ref{fig:spacing_window}(a) shows that the leading formula follows both alternating spacing subsequences.
Panel (b) shows that Eq.~(\ref{snapprox}) is more accurate than the leading result in Eq.~(\ref{spacingLeading}).
The maximum error of the leading result is $7.593\times10^{-4}$, below two percent of the local envelope amplitude.
Appendix~\ref{app:asymptotic_accuracy} extends this comparison to successive index windows.
For the full benchmark window $[8000,10^5)$, the maximum leading-spacing error is $8.37\times10^{-4}$, and its RMS value is $2.15\times10^{-4}$.
The maximum error relative to the local spacing envelope is $1.99\%$.
These tests assess the asymptotic approximation at the stated parameters, separately from basis-cutoff convergence. 
They do not infer a parameter-independent onset index.

We consider the spacing indices $n=m_-,\ldots,m_+-1$ in the window $n\in[m_-,m_+)$.
We select $m_-$ by requiring Eq.~(\ref{asymptoticEigenvalues}) to be accurate throughout the window for the specified $g$ and $\Delta$. Appendix~\ref{app:asymptotic_accuracy} quantifies this accuracy for the central benchmark.
The condition $r m_-^{-1/4}<1$ keeps the leading spacings positive, and $2g/\sqrt{m_-}\ll1$ ensures a small phase change between adjacent indices.
For the envelope $A(n)=r n^{-1/4}$, one phase period has index length $T_n\simeq\pi\sqrt n/g$.
Its relative change over that period is $|A'(n)|T_n/A(n)\simeq\pi/(4g\sqrt n)$.
We require this quantity to be small and approximate $A(n)$ by its value at the center of each local period, while retaining its slow variation across the full window.
Finally, the continuum count requires the window to sample many oscillations:
\begin{equation}
N_{\mathrm{osc}}=\frac{2g}{\pi}\left(\sqrt{m_+}-\sqrt{m_-}\right)\gg1.
\end{equation}
Spectral unfolding maps an energy $E$ to the smooth integrated level count $\overline{N}(E)$, removing the smooth variation of the mean density \cite{Guhr_1998}.
The unit mean spacing of the leading ladder allows us to use the raw dimensionless spacings without a further unfolding transformation.
The histograms below are normalized separately to unit area.

\subsection{Counting formula}

\begin{figure}[!t]
\centering
	\includegraphics[width=\linewidth]{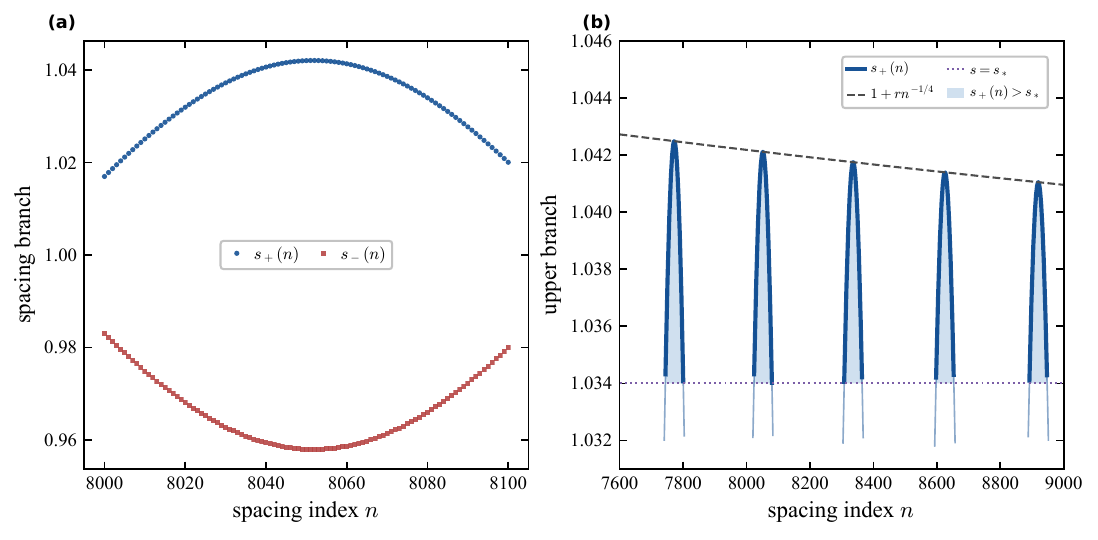}
	\caption{Analytic spacing branches and envelope for $g=\Delta=1$.
	(a) Integer values of the plus branch $s_+(n)$ (blue circles) and minus branch $s_-(n)$ (red squares) in the window $n\in[8000,8100]$.
	(b) Portions of $s_+(n)$ near its local maxima over $n\in[7600,9000]$.
	The blue curve is $s_+(n)$, the gray dashed curve is the upper envelope $1+rn^{-1/4}$, the purple dotted line is a representative threshold $s=s_*$, and the pale blue regions mark the intervals for which $s_+(n)>s_*$.}
	\label{fig:branches_envelope}
\end{figure}

The two spacing branches are reflected about $s=1$, as shown in Fig.~\ref{fig:branches_envelope}(a).
The shaded intervals in panel (b) mark where $s_+(n)$ exceeds a chosen illustrative threshold $s_*=1.034$ near each displayed local maximum.
Below, $s$ denotes a variable threshold, $s_*$ is only its fixed value in the figure.
As $n$ increases, the intervals narrow with the decaying envelope and cease to contribute once the threshold can no longer be reached.

The largest oscillation amplitude occurs at the lower end of the window and sets the leading continuum support $[s_{\min},s_{\max}]$, with
\begin{equation}\label{spacingSupport}
s_{\min}=1-rm_-^{-1/4},\qquad s_{\max}=1+rm_-^{-1/4}.
\end{equation}
The window contains $m_+-m_-$ spacings.
For $1<s<s_{\max}$, define the discrete exceedance count by
\[
\mathcal{N}(s)=\#\{n\in\mathbb{Z}:m_-\leq n<m_+,\ s(n)>s\}.
\]
Thus $s(n)$ is an individual spacing and $s$ is the threshold against which all spacings are compared.
In the leading continuum approximation, the lower tail follows by reflection about $s=1$.

Consider first the plus branch.
Its continuous local maxima lie near
\begin{equation}\label{nj}
n_j=\left(\frac{\pi(2j+1/4)}{a}\right)^2.
\end{equation}
Rounding to the appropriate sampled index changes $n_j$ only by $O(1)$, which is negligible when $2g/\sqrt{m_-}\ll1$.
Over the local oscillation centered near $n_j$, we approximate the amplitude by $A_j=r n_j^{-1/4}$ and retain the phase dependence on $n$.
The two threshold-crossing indices around that maximum, $n_j^-$ and $n_j^+$, solve $s_+(n_j^\pm)=s$.
In the local fixed-envelope approximation, we solve $1+A_j\cos(a\sqrt n-\pi/4)=s$ and obtain
\begin{equation}
a\sqrt{n_j^\pm}-\frac{\pi}{4}\simeq2\pi j\pm\phi_j,
\qquad
\phi_j=\arccos\left(\frac{s-1}{r}n_j^{1/4}\right).
\end{equation}
The corresponding interval width is
\begin{equation}
n_j^+-n_j^-\simeq\frac{4\sqrt{n_j}}{a}
\arccos\left(\frac{s-1}{r}n_j^{1/4}\right).
\end{equation}
Since $dn_j/dj\simeq4\pi\sqrt{n_j}/a$, one family of maxima has density $a/(4\pi\sqrt n)$.
The two branches provide two families of maxima shifted by half a phase period.
Within a fixed even-parity sector, only odd indices sample $s_+$ and only even indices sample $s_-$.
Consequently, an interval of length $L$ in either branch contributes approximately $L/2$ sampled spacings.
Adding the two families multiplies the count by two, giving the combined weight $2\times(1/2)=1$ before integration over the maxima.
Only maxima below the envelope cutoff
\begin{equation}
n_{\max}(s)=\left(\frac{r}{s-1}\right)^4
\end{equation}
can exceed the threshold $s$.
Combining the interval width and maxima density gives
\begin{equation}\label{ns}
\mathcal{N}(s)\simeq\frac{1}{\pi}
\int_{m_-}^{\min\{m_+,n_{\max}(s)\}}dn\,
\arccos\left(\frac{s-1}{r}n^{1/4}\right),
\qquad 1<s<s_{\max}.
\end{equation}
The leading continuum count vanishes for $s\geq s_{\max}$.
Taking $s\to1^+$ in Eq.~(\ref{ns}) gives $\mathcal{N}(1^+)\simeq(m_+-m_-)/2$, as required by reflection symmetry.

\subsection{Analytic density and twin peaks}

Let $P(s)$ denote the normalized spacing density obtained by differentiating the leading continuum count in Eq.~(\ref{ns}).
It vanishes outside the support in Eq.~(\ref{spacingSupport}) and satisfies
\begin{equation}
\int_{s_{\min}}^{s_{\max}}P(s)\,ds=1.
\end{equation}
For $1<s<s_{\max}$, the count above $s$ is
\begin{equation}
\mathcal{N}(s)\simeq(m_+-m_-)\int_s^{s_{\max}}P(s')\,ds'.
\end{equation}
Differentiating the leading continuum count in Eq.~(\ref{ns}) yields the $s>1$ branch of the spacing density,
\begin{equation}\label{density}
P(s)=\frac{1}{m_+-m_-}\frac{1}{\pi r}
\int_{m_-}^{\min\{m_+,n_{\max}(s)\}}dn\,
\frac{n^{1/4}}{\sqrt{1-\left(\frac{s-1}{r}n^{1/4}\right)^2}}.
\end{equation}
Since $s_-(n)=2-s_+(n)$, the full density obeys
\begin{equation}\label{symmetry}
P(s)=P(2-s).
\end{equation}
As a normalization check, Eq.~(\ref{density}) integrates to one half over $s>1$, and the reflection in Eq.~(\ref{symmetry}) supplies the other half.
Together with the zero density outside the support, these relations determine the full normalized distribution.

The square-root kernel in Eq.~(\ref{density}) connects the finite-window density to the arcsine law in Theorem~1.4 of Ref.~\cite{Rudnick2024}.
That theorem concerns the rescaled eigenvalue deviations $n^{1/4}[\lambda^p(n)-(n-g^2)]$, whose oscillation amplitude in our convention is $r/2$. The parameter multiplying $\sigma_z$ in Ref.~\cite{Rudnick2024} is $\Delta/2$ here.
For the spacings, the same oscillatory phase produces a local arcsine density with amplitude $A(n)=rn^{-1/4}$.
Under the slow-envelope and many-oscillation conditions stated above, Eq.~(\ref{density}) and its reflection can therefore be written as the mixture
\[
P(s)=\frac{1}{m_+-m_-}\int_{m_-}^{m_+}
\frac{\mathbf{1}_{\{|s-1|<A(n)\}}}{\pi\sqrt{A(n)^2-(s-1)^2}}\,dn,
\qquad A(n)=rn^{-1/4},
\]
where the integrand is zero when $|s-1|\geq A(n)$.
Each local kernel has unit mass, and retaining its changing amplitude across the window gives the support edges and cusps discussed below.
The endpoints $\alpha,\beta$ in the cited theorem bound values of the rescaled deviation. The index endpoints $m_\pm$ enter here through the envelope amplitudes $A(m_\pm)$.

The two square-root cusps occur at
\begin{equation}
s=1\pm r m_+^{-1/4}.
\end{equation}
At the upper cusp, $n_{\max}(s)=m_+$ and the integration endpoint changes from $m_+$ to $n_{\max}(s)$.
This endpoint switch combines with the square-root denominator in Eq.~(\ref{density}) to produce the cusp, reflection gives the lower one.
Thus $m_-$ controls the support, and $m_+$ determines the cusp positions.

\begin{figure}[!t]
\centering
	\includegraphics[width=.88\linewidth]{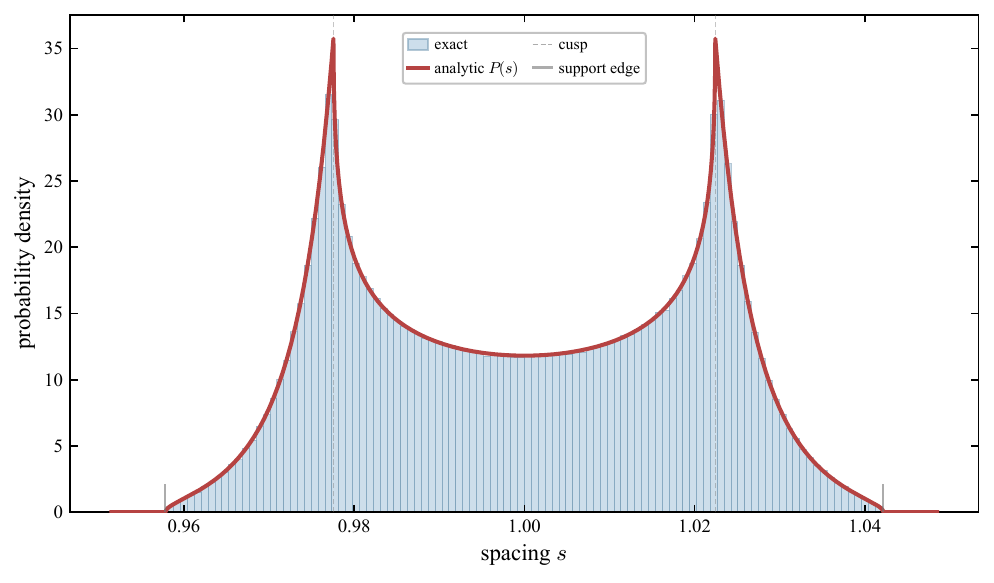}
	\caption{Spacing distribution benchmark for $g=\Delta=1$ in the window $n\in[m_-,m_+)$, with $m_-=8000$ and $m_+=10^5$.
	The blue histogram contains spacings from exact diagonalization in the even-parity sector, using 120 bins without smoothing. The red curve is the full leading density obtained from Eq.~(\ref{density}) for $s>1$ and Eq.~(\ref{symmetry}) for $s<1$ with the same window parameters.
	Dashed lines mark the leading cusp positions $s=1\pm r m_+^{-1/4}$. Short edge ticks mark the leading support $s=1\pm r m_-^{-1/4}$.
	Each dashed line lies within one histogram bin of the corresponding numerical maximum.
	The edge ticks show the unbinned leading support, the outer occupied bins straddle these predicted edges.}
	\label{fig:spacing_histogram_analytic}
\end{figure}

Figure~\ref{fig:spacing_histogram_analytic} shows that the analytic density reproduces the numerical histogram across the full support, including the twin peaks.
The oscillation amplitude decreases across the index window: its larger value at $m_-$ sets the outer support edges, and its smaller value at $m_+$ sets the two cusp positions.
The observed edges and peak positions agree with these predictions at the histogram resolution.

\begin{figure}[!t]
\centering
	\includegraphics[width=\linewidth]{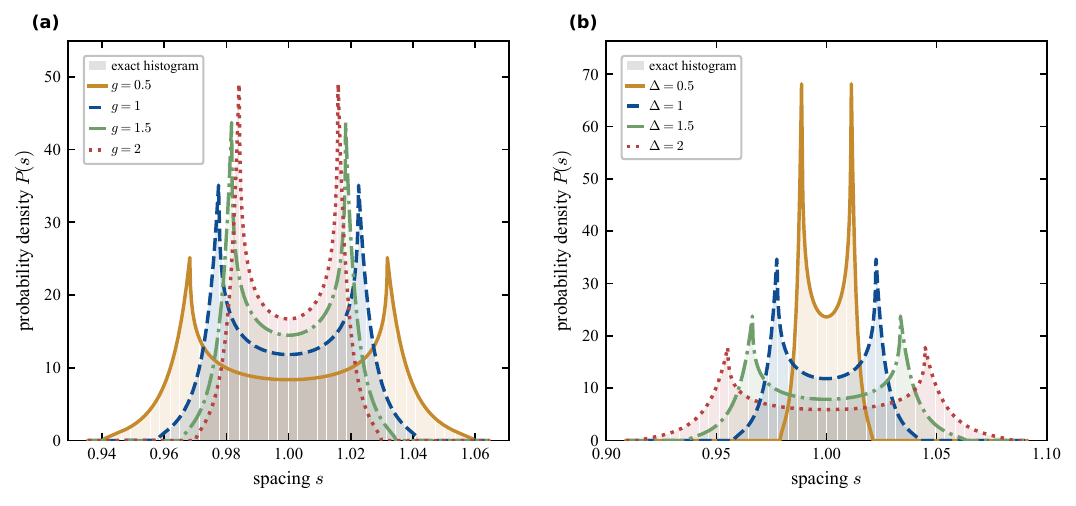}
\caption{Parameter dependence of the spacing density in the even-parity sector for the same window $n\in[m_-,m_+)$, with $m_-=8000$ and $m_+=10^5$.
	(a) Fixed $\Delta=1$ with $g=0.5,1,1.5,2$.
	(b) Fixed $g=1$ with $\Delta=0.5,1,1.5,2$.
		In each panel, the orange solid, blue dashed, green dash-dotted, and red dotted lines correspond to the four parameter values in increasing order.
		For each parameter value, the lightly shaded histogram and analytic line use the same color, the histograms are obtained from exact diagonalization with 180 bins, and the lines are the leading densities obtained from Eqs.~(\ref{density}) and~(\ref{symmetry}).
	The motion of the support and cusp positions follows the amplitude $r=\Delta/\sqrt{2\pi g}$.}
	\label{fig:spacing_robustness}
\end{figure}

The twin peaks move with the amplitude coefficient $r=\Delta/\sqrt{2\pi g}$, as shown in Fig.~\ref{fig:spacing_robustness}.
At fixed $\Delta$, increasing $g$ decreases $r$ and moves the cusps toward $s=1$ in panel (a), at fixed $g$, increasing $\Delta$ increases $r$ and moves them outward in panel (b).
Across both panels, the analytic curves follow the same motion in the histograms from exact diagonalization.

These trends agree with the known oscillator limits.
At $\Delta=0$, the displaced ladders in Sec.~II.B give $s(n)=1$ exactly and hence $P(s)=\delta(s-1)$.
The narrowing with increasing $g$ is consistent with the strong-coupling limit $\lambda^p(n)+g^2\to n$ for each fixed $n$ in either parity sector \cite{HiroshimaShirai2025}.
This strong-coupling result is an independent limit, the present high-energy expansion is taken at fixed $g$.
At fixed $m_-$, taking $m_+\to\infty$ gives weak convergence to $\delta(s-1)$ \cite{ThiHoaiNguyen2024}: the cusps approach unity and the normalized weight away from unity vanishes, while the finite-window support edges remain fixed.
Moving both endpoints upward also makes the support itself shrink, provided the window continues to sample many oscillations.

\section{Ratio of consecutive level spacings}

The spacing density describes the distribution of individual gaps, whereas the ratio of consecutive spacings also probes how neighboring gaps are paired.
For the QRM, the slowly varying oscillation and the alternating factor in Eq.~(\ref{spacingLeading}) predict opposite deviations of neighboring spacings from the unit ladder.
The ratio therefore measures the size of this local alternation and its decay at high energy.
This statistic requires no spectral unfolding because a common local rescaling of the two gaps cancels in their ratio \cite{Oganesyan2007,Atas_2013}.
Within one parity sector, we define the ratio directly from adjacent spacings as
\begin{equation}\label{fixedParityRatio}
\rho_n=\frac{\min\{s(n-1),s(n)\}}{\max\{s(n-1),s(n)\}},
\qquad 0<\rho_n\leq1.
\end{equation}

Between adjacent indices, the envelope and phase in Eq.~(\ref{spacingLeading}) vary slowly, whereas the alternating factor changes sign.
Let $u_n$ denote the leading magnitude of the deviation from unit spacing.
To leading order, the two spacings entering $\rho_n$ are $1-u_n$ and $1+u_n$, in either order.
Hence
\begin{equation}\label{leadingRatio}
u_n\equiv r n^{-1/4}\left|\cos\left(a\sqrt n-\frac{\pi}{4}\right)\right|,
\qquad
\rho_n^{(0)}=\frac{1-u_n}{1+u_n}.
\end{equation}
The superscript $(0)$ denotes the leading asymptotic approximation: the spectral remainder and the variation of the envelope and phase between the two adjacent spacings are omitted.
The positivity condition $r m_-^{-1/4}<1$ ensures that both leading spacings are positive throughout the window.
For spacing indices $n\in[m_-,m_+)$, the ratio indices are $m_-+1,\ldots,m_+-1$.

The amplitude at the lower end of the index window, $r m_-^{-1/4}$, sets the leading support $\rho\in[\rho_{\min},1]$.
The two spacing cusps have the same deviation magnitude $u_*=r m_+^{-1/4}$ and therefore map to the same ratio through Eq.~(\ref{leadingRatio}).
We call this value the characteristic ratio $\rho_*$ of the finite QRM window:
\begin{equation}\label{ratioScales}
\rho_{\min}=\frac{1-rm_-^{-1/4}}{1+rm_-^{-1/4}},
\qquad
\rho_*\equiv\frac{1-rm_+^{-1/4}}{1+rm_+^{-1/4}}.
\end{equation}
Thus $\rho_*$ is a characteristic scale determined by the QRM parameters and the upper window endpoint.
At fixed $g$ and $\Delta$, shifting the window to higher energies moves both $\rho_{\min}$ and the characteristic ratio $\rho_*$ toward unity, as long as the window continues to sample many oscillations.
The concentration near unity reflects the approach to equal neighboring gaps, while their finite deviation from unity retains the oscillatory structure of the QRM spectrum.

\begin{figure}[!t]
\centering
	\includegraphics[width=.94\linewidth]{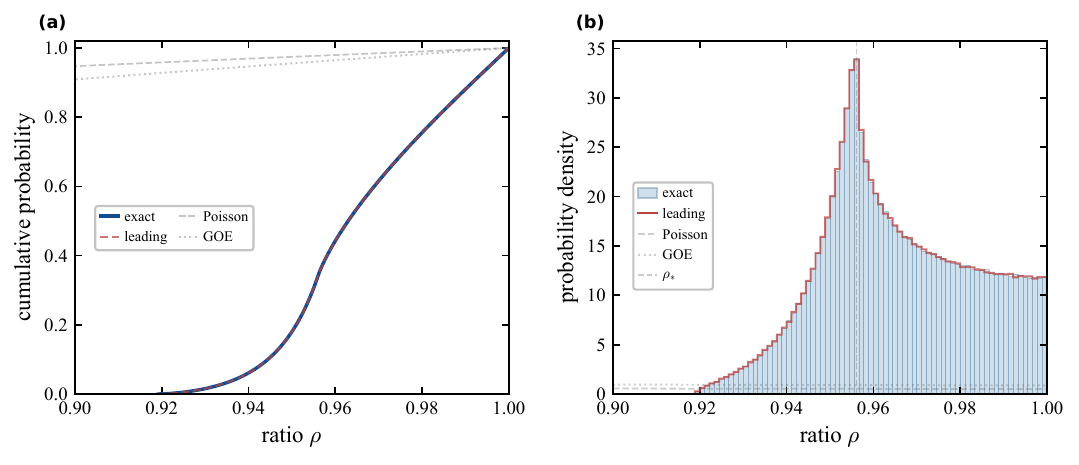}
	\caption{Ratio of consecutive level spacings in the even-parity sector for $g=\Delta=1$, using spacing indices $n\in[8000,10^5)$ and ratio indices $n=8001,\ldots,99999$.
	(a) Empirical cumulative distributions from exact diagonalization (blue solid curve) and Eq.~(\ref{leadingRatio}) (red dashed curve). Pale gray dashed and dotted curves show the Poisson reference and GOE surmise.
	(b) Densities near $\rho=1$ from exact diagonalization (blue histogram) and Eq.~(\ref{leadingRatio}) (red step curve), using 90 bins over the displayed interval and the same gray references.
	The vertical gray line marks the characteristic ratio $\rho_*=0.95612$ from Eq.~(\ref{ratioScales}).}
	\label{fig:ratio_benchmark}
\end{figure}

To compare the full ratio distributions, we use the cumulative distribution function $F(\rho)=\Pr(\rho_n\leq\rho)$.
Figure~\ref{fig:ratio_benchmark} compares the leading ratio statistics with exact diagonalization and with the Poisson and GOE references.

For the central benchmark and window used above, increasing $N_{\mathrm{cut}}$ from $110000$ to $120000$ changes the empirical ratio distribution by a Kolmogorov--Smirnov distance of $2.2\times10^{-5}$.
The Poisson reference density and the standard GOE surmise are \cite{Atas_2013}
\begin{equation}\label{ratioReferences}
P_{\rm P}(\rho)=\frac{2}{(1+\rho)^2},
\qquad
P_{\rm GOE}(\rho)=\frac{27}{4}\frac{\rho+\rho^2}{(1+\rho+\rho^2)^{5/2}}.
\end{equation}
The latter is obtained from three-dimensional GOE matrices and approximates the bulk ratio distribution for large matrices.

In panel (a), the exact and leading cumulative distributions nearly coincide over the displayed range, 
with a Kolmogorov-Smirnov distance of $7.2\times10^{-4}$.
In panel (b), the exact and leading densities attain their maximum in the same bin, which contains the predicted value $\rho_*=0.95612$.
The enhancement near $\rho_*$ is the ratio counterpart of the spacing cusps under the map in Eq.~(\ref{leadingRatio}).
The much narrower QRM distribution compared with the smooth Poisson and GOE references records the near-uniform ladder and its correlated alternating modulation.

The counting method requires a controlled spectral asymptotic form, spacing branches that can be counted separately and have slowly varying amplitudes, and a window containing many oscillations.
Applications to another model require establishing these ingredients for its own spectrum.
Related large-eigenvalue asymptotics have been obtained for the asymmetric and two-photon quantum Rabi models \cite{Charif2024,BoutetDeMonvel2024TwoPhoton}, providing concrete settings in which the applicability of this construction can be explored.

\section{Conclusion}

In summary, we have derived an analytic expression for the spacing density $P(s)$ in a fixed parity sector of the quantum Rabi model.
At high energies, the spectrum approaches a uniformly spaced ladder modulated by an oscillation whose amplitude decays as $n^{-1/4}$.
The alternating sign of this oscillation separates consecutive spacings into two branches.

For a finite index window containing many oscillations, we obtain $P(s)$ at leading order by counting the branch intervals above each spacing threshold and differentiating the cumulative count.
The oscillation amplitude at the lower end of the window sets the support of $P(s)$, and its value at the upper end determines the positions of the two square-root cusps that form the twin peaks.
Spacing histograms from exact diagonalization agree with the analytic density, including the predicted support and peak positions.

At leading order, the same spectral analysis yields an analytic expression for the ratio $\rho_n$ of consecutive spacings, a statistic that requires no spectral unfolding.
The oscillation amplitude at the upper end of the window determines the characteristic ratio $\rho_*$, near which the ratio density is enhanced.
The resulting ratio statistics agree closely with those obtained from exact diagonalization.

As an index window that continues to contain many oscillations moves to higher energies, the decaying oscillation brings the two spacing branches closer together.
Accordingly, the twin peaks of $P(s)$ and the characteristic ratio $\rho_*$ both move toward unity.
Over a finite window, the twin peaks and the enhancement near $\rho_*$ reveal the decaying spectral oscillation even as the individual spacings approach unity.
The same counting approach can be explored in other systems with controlled spectral asymptotics, slowly varying spacing branches, and windows containing many oscillations.
The present leading continuum analysis concerns high-energy windows containing many oscillations. Low-energy statistics and finer structures beyond this approximation remain topics for further study.

\begin{acknowledgments}
    This work is supported by the National Natural Science Foundation of China Grant No. 12205383, Hunan Provincial Natural Science Foundation of China No. 2024JJ6483, and Australian Research Council Grant No. DP210102243.
    G.S. acknowledges financial support from the Spanish MINECO grant PID2024-161474NB-I00 (MCIU/AEI/FEDER, UE), and from QUITEMAD-CM TEC2024/COM-84.
\end{acknowledgments}

\appendix
\section{Numerical accuracy of the asymptotic spectrum}
\label{app:asymptotic_accuracy}
We quantify the asymptotic approximation at fixed $g=\Delta=1$ in the even-parity sector.
Let $\lambda_n^{\mathrm{as}}$ denote the displayed leading expression in Eq.~(\ref{asymptoticEigenvalues}), and let $s_n^{(0)}$ denote Eq.~(\ref{spacingLeading}).
For $m_-\leq n<m_+$, define
\[
 e_n=\lambda(n)-\lambda_n^{\mathrm{as}},\qquad
 d_n=[\lambda(n+1)-\lambda(n)]-s_n^{(0)}.
\]
The numerical eigenvalues use $N_{\mathrm{cut}}=110000$.
Table~\ref{tab:asymptotic_accuracy} shows decreasing maximum and root-mean-square (RMS) errors over the successive windows tested.
For the manuscript window $[8000,10^5)$, the maximum of $|d_n|/(r n^{-1/4})$ is $1.99\%$, and its RMS value is $0.679\%$.
This normalization uses the oscillation envelope and remains finite where the cosine vanishes.
The corresponding maximum eigenvalue error relative to its envelope $(r/2)n^{-1/4}$ is $0.892\%$.
These are empirical checks at the stated fixed parameters and do not provide a parameter-uniform lower bound on $m_-$ or imply pointwise monotonic convergence.

\begin{table}[!t]
\caption{Errors of the asymptotic eigenvalues and leading spacings for $g=\Delta=1$ in the even-parity sector.
All maxima and RMS values are taken over the spacing indices $n\in[m_-,m_+)$. Evaluating the final spacing requires the eigenvalue at $m_+$.
$N_{\mathrm{osc}}=(2g/\pi)(\sqrt{m_+}-\sqrt{m_-})$ is the phase coverage in oscillations.
The final row gives the window used in the central distribution benchmark.}
\label{tab:asymptotic_accuracy}
\small
\begin{tabular}{rrrrr}
\hline\hline
$[m_-,m_+)$ & $N_{\mathrm{osc}}$ & $\max|e_n|$ & $\max|d_n|$ & $\mathrm{RMS}(d_n)$ \\
\hline
$[1000,2000)$ & 8.34 & $8.21\times10^{-4}$ & $3.81\times10^{-3}$ & $2.12\times10^{-3}$ \\
$[2000,4000)$ & 11.79 & $4.99\times10^{-4}$ & $2.29\times10^{-3}$ & $1.26\times10^{-3}$ \\
$[4000,8000)$ & 16.68 & $3.10\times10^{-4}$ & $1.40\times10^{-3}$ & $7.68\times10^{-4}$ \\
$[8000,16000)$ & 23.59 & $1.88\times10^{-4}$ & $8.37\times10^{-4}$ & $4.60\times10^{-4}$ \\
$[16000,32000)$ & 33.36 & $1.15\times10^{-4}$ & $5.06\times10^{-4}$ & $2.77\times10^{-4}$ \\
$[32000,64000)$ & 47.17 & $7.00\times10^{-5}$ & $3.05\times10^{-4}$ & $1.67\times10^{-4}$ \\
$[64000,100000)$ & 40.26 & $4.28\times10^{-5}$ & $1.84\times10^{-4}$ & $1.10\times10^{-4}$ \\
\hline
$[8000,100000)$ & 144.38 & $1.88\times10^{-4}$ & $8.37\times10^{-4}$ & $2.15\times10^{-4}$ \\
\hline\hline
\end{tabular}
\end{table}

As a separate check of numerical accuracy, we recomputed adjacent eigenvalue pairs beginning at $n=1000,1999,8000,15999,32000,63999,99999$ with $N_{\mathrm{cut}}=120000$ using a selective bisection eigensolver.
The maximum spacing difference among these sampled pairs is $1.02\times10^{-10}$, far below the asymptotic errors in Table~\ref{tab:asymptotic_accuracy}.
This basis-size and eigensolver check addresses numerical convergence, whereas the table addresses the accuracy of the asymptotic approximation as the index increases.

\renewcommand{\baselinestretch}{1.39}\selectfont
\bibliography{LevelStatisticsQRM_revision}

\end{document}